\documentclass[a4paper,11pt]{article}
\usepackage{pos}
\usepackage{multirow}
\usepackage{footnotehyper}
\makesavenoteenv{table}
\makesavenoteenv{tabular}

\title{Can Historic Measurements cast light on the Spectral Differences found between the Pierre Auger Observatory and the Telescope Array?}
\ShortTitle{Historic Measurements}

\author[a]{Quentin Luce}
\author[b]{Alan A. Watson}

\affiliation[a]{IJCLab, Universit\'{e} Paris-Saclay, CNRS/IN2P3, Orsay, France}
\affiliation[b]{School of Physics and Astronomy, University of Leeds, Leeds, United Kingdom}

\emailAdd{lucequentin.pro@gmail.com}
\emailAdd{a.a.watson@leeds.ac.uk}

\abstract{The differences between the energy spectra reported by the Pierre Auger Observatory and the Telescope Array Collaboration are well-known.  In an effort to understand these differences, we have studied the residuals of the fluxes measured in early work at AGASA, Volcano Ranch and Haverah Park with reference to the spectrum reported by the Auger Collaboration.  After an empirical adjustment to the energies in the region before the ankle, where fluxes from the two hemispheres are expected to be the same, it is found that historic data are in good agreement with measurements made at the Pierre Auger Observatory above 30 EeV.  Thus, the tensions between the spectra measured by the Telescope Array and the Auger Observatory remain.}

\FullConference{29th European Cosmic Ray Symposium\\
29 June – 3 July 2026\\
Santiago de Compostela, Spain\\
}

\begin{document}
\maketitle

\section{Introduction}

The differences between the energy spectra reported by the Pierre Auger Observatory and the Telescope Array Collaborations have been quantified and documented over the last decade, and are found even when fluxes in the declination band observed by both instruments are compared~\cite{WGSpectrum_ICRC2025}.
From an exposure of $\sim$$1.4 \times 10^4\,\mathrm{km}^2\mathrm{sr}\,\mathrm{yr}$, largely from the Northern Hemisphere, the Telescope Array Collaboration have reported 18 events above 100 EeV for an integral flux of $(1.3 \pm 0.3) \times 10^{-3}\,\mathrm{km}^{-2}\,\mathrm{sr}^{-1}\,\mathrm{yr}^{-1}$~\cite{TASpectrum_ICRC2025}.
The corresponding measurement from the Pierre Auger Observatory~\cite{AugerSpectrum_PRD2020, AugerSpectrum_PRL2025}, based largely on observations from the Southern Hemisphere and from an exposure of $6 \times 10^{4}\,\mathrm{km}^2\,\mathrm{sr}\,\mathrm{yr}$, is about a factor of 5 lower at $(2.5 \pm 0.7) \times 10^{-4}\,\mathrm{km}^{-2}\,\mathrm{sr}^{-1}\,\mathrm{yr}^{-1}$, with 24 events reported above 100 EeV.
In the second half of the $20^\mathrm{th}$ Century, 16 events were reported from the Volcano Ranch, Haverah Park and AGASA arrays as having energies above 100 EeV, giving a much greater integral flux of $(8 \pm 2) \times 10^{-3}\,\mathrm{km}^{-2}\,\mathrm{sr}^{-1}\,\mathrm{yr}^{-1}$ from a combined exposure about seven times smaller than Telescope Array and $\sim$30 times less than from the Auger Observatory.

The spectra reported by the Auger Collaboration~\cite{AugerSpectrum_PRD2020, AugerSpectrum_PRL2025} are data-driven, with no assumptions being made about the primary mass or features of hadronic interactions at the relevant energies. 
We take these measurements as a reference in the analysis bellow.
By contrast, for energy estimates from the ground-arrays operated in the Northern Hemisphere, assumptions were made about shower development (Volcano Ranch) or by using mass-dependent models rooted in pre-LHC data (Haverah Park, AGASA).
The approach of the Telescope Array Collaboration is following this philosophy using models updated with post-LHC data.
To update the energy estimates from the earlier instruments, the historic events have been re-evaluated using the EPOS-LHC~\cite{Pierog_EPOS-LHC} model of hadronic interactions (Section 2).
With this model, a significant disagreement with the spectrum from the Auger Observatory remains, even under the most favourable assumption of proton primaries.

Consequently, another approach has been explored by examining the residuals of the fluxes with reference to the spectrum from the Auger Observatory.
After an empirical adjustment in the energy region before the ankle where fluxes from the two hemispheres are surely the same, good agreement is obtained (Section 3).
However, when these adjustments are extended to higher energies, it is found that the flux measurement above 30 EeV from the AGASA array, from which most of the historic exposure comes, is in good accord with that from the Pierre Auger Observatory.
Although agreement between the Auger spectrum and the historic measurements is found, the tensions with the Telescope Array result remain (Section 4).

\section{Results from an updated shower model}

At energies of $10^{18.5}$, $10^{19.0}$, $10^{19.5}$, and $10^{20.0}$ eV and zenith angles of 0, 30 and 45$^{\circ}$, simulations tailored to Volcano Ranch, Haverah Park and AGASA have been carried out using the EPOS-LHC model.
For the Volcano Ranch and Haverah Park arrays, detailed information is available for a relatively large number of events claimed to have energies above 10 EeV~\cite{YellowBook}, while for AGASA four very energetic events have been discussed in detail~\cite{Hayashida_1994, Doi_1997, Sakaki_2001}. 
This information is of crucial importance for re-analysis of events at the detector level, prior to any assumptions made during the reconstruction procedure. 
Using the simulations, the revised energy estimate is $E_\mathrm{mod} = kE$, where $E$ is the original estimate of the energy.
Values of $k$ computed for Volcano Ranch, Haverah Park and AGASA, assuming proton and iron primaries (p, Fe) are (0.72, 0.71), (0.73, 0.65) and (0.66, 0.59) respectively.
The uncertainties in these numbers are $<6\%$ for proton and $<2\%$ for iron.
If these estimates are correct, then the energy of the famous 100 EeV event reported in 1963 at Volcano Ranch~\cite{Linsley_1963} would be reduced to $\sim$70 EeV.
The revised Haverah Park energies align with the re-analysis previously carried out for lower energies in \cite{Ave_2003}. 
\begin{figure}[t]
    \centering
    \includegraphics[width=0.57\columnwidth]{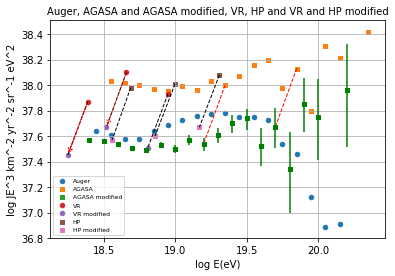}
    \caption{$\log E^3J$ vs $\log E$ for the AGASA, Volcano Ranch and Haverah Park as originally reported. The AGASA spectrum with the energies reduced by the factor $k$ discussed in the text, are shown in green.}
    \label{fig:corrected_spectra}
\end{figure}

The effect of these corrections on data from AGASA, Volcano Ranch and Haverah Park is shown in Figure~\ref{fig:corrected_spectra} where they are displayed using the traditional $\log E^3J$ vs $\log E$ plot.
The Auger and AGASA spectra are in poor agreement, as are the sample of measurements shown from Volcano Ranch and Haverah Park, when assuming an energy/zenith-independent factor $k$.
The corrections to the historic data have been made under the assumption that the primary particles are protons, although arguments, independent of assumptions about hadronic interactions, appear to invalidate this assumption for data from the Northern Hemisphere~\cite{Watson_2026}.
Adopting a heavier mean mass would enhance the disagreement.
While the AGASA data points agree with those from Volcano Ranch and Haverah Park, both before and after the correction, all data, except for those based on the 11 highest-energy AGASA events, fall below the Auger spectrum.
These 11 events gave rise to much debate with proposals to explain them, including the breakdown of Lorentz invariance~\cite{Coleman_1999} and that the particles might be evidence for cosmological relics~\cite{Berezinsky_2000}.
The accuracy of the energy estimates of these events is discussed below (Section 4).

\section{Approach using residuals from a reference spectrum}
\begin{figure}[t]
    \centering
    \includegraphics[width=0.48\columnwidth]{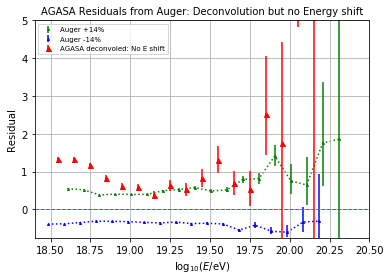}
    \includegraphics[width=0.5\columnwidth]{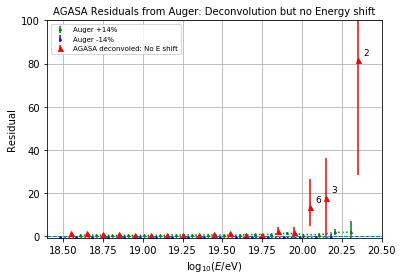}
    \caption{Residuals of the AGASA spectrum with respect to the Auger spectrum after deconvolution but with no adjustment of the energy estimate. The two plots have different scales on the y-axes to better highlight the discrepancies in the different energy ranges.}
    \label{fig:AGASA_residuals}
\end{figure}

We define the residual from a given spectrum as Residual = ($J/J_{s}$ – 1), where $J$ is the flux measured in some energy interval and $J_s$ is the flux for an arbitrary reference spectrum, taken here as the approximation to the spectrum reported by the Auger Collaboration~\cite{AugerSpectrum_PRD2020},
\begin{equation}
    \label{eq:auger_spectrum}
    J(E) = J_0 \left(\frac{E}{E_0}\right)^{-\gamma_1}\prod_{i=1}^3\left[ 1 + \left(\frac{E}{E_{ij}}\right)^{\frac{1}{\omega_{ij}}} \right]^{(\gamma_i-\gamma_j)\omega_{ij}},
\end{equation}
a spectrum with four slopes and three breaks.
We make the following assumptions:
\begin{itemize}
    \itemsep0em
    \item The estimate of the energy spectrum from the Auger Collaboration above 3 EeV is the best available, with a systematic uncertainty in the measurement of $\pm 14\%$~\cite{AugerSpectrum_PRD2020}.
    \item At energies lower than the ankle ($\sim$5 EeV) the fluxes from both hemispheres are the same, an assumption supported by the negligible anisotropy in arrival directions~\cite{Deligny_ECRS2026} and the independence of mass composition on declination~\cite{AugerXmax_PRD2026}.
\end{itemize}
In treating the AGASA data, the effect of the energy uncertainty on the measured flux at a given energy has been accounted for with an approximate unfolding of the spectrum using the well-known relation for the ratio of the observed to the true number of events:
\begin{equation}
    \label{eq:unfolding}
    \exp\left[ \frac{(\gamma - 1)^2 \sigma^2_{\ln E}}{2}\right],
\end{equation}
where the differential slope, $\gamma$, and the uncertainty, $\sigma$, are taken from~\cite{Takeda_2003}.
Typically, the number of events is reduced by a factor between 0.88 and 0.94, depending on the energy interval.
The residuals for the AGASA spectrum with unfolding, but with no adjustment to the energy scale, are shown in Figure~\ref{fig:AGASA_residuals}.
The disagreement with the Auger measurement~\cite{AugerSpectrum_PRD2020} is evident.

\begin{figure}[t]
    \centering
    \includegraphics[width=0.49\columnwidth]{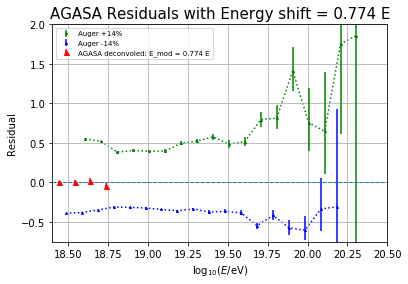}
    \includegraphics[width=0.49\columnwidth]{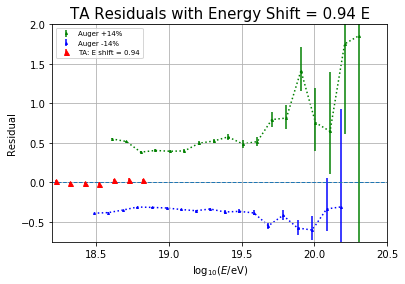}
    \caption{Residual plots for the data from AGASA (left) and Telescope Array (right) at energies below the energy of the ankle. The reduced values of $\chi^2$ are 0.35 and 1.7 respectively.}
    \label{fig:AGASA_TA_residuals_lowenergy}
\end{figure}

In Figure~\ref{fig:AGASA_TA_residuals_lowenergy}, residual plots are shown for AGASA ~\cite{Takeda_2003} and Telescope Array~\cite{TASpectrum_ICRC2025} data but only for measurements at energies below the ankle ($< 5$EeV).
Using a $\chi^2$-minimisation, the shifts in energy for best alignment are $(0.77 \pm 0.01)$ for AGASA and $(0.94 \pm 0.01)$ for Telescope Array.
Note that the optimal shift for the AGASA data is smaller than that found for a pure proton beam ($k = 0.66$) using the EPOS-LHC model and that there is evidence that protons, He and N dominate in this energy region~\cite{AugerXmax_PRD2026}.
The small adjustment required to bring the Telescope Array measurement into accord with the Auger result is consistent with that reported in \cite{WGSpectrum_ICRC2025}, thus validating the residual approach.

\section{Comparison of historic data with Auger and Telescope Array measurements}

\begin{figure}[t]
    \centering
    \includegraphics[width=0.49\columnwidth]{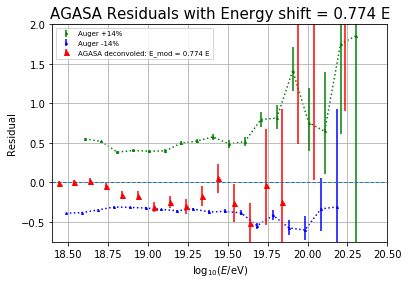}
    \includegraphics[width=0.49\columnwidth]{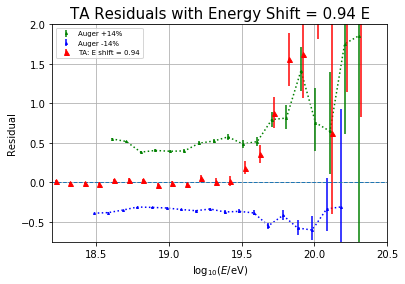}
    \caption{Residuals plots for the whole energy range measured at the AGASA (left) and Telescope Array (right) experiments. The reduced values of $\chi^2$ are 4.35 and 7.01 respectively.}
    \label{fig:AGASA_TA_residuals}
\end{figure}
\begin{figure}[t]
    \centering
    \begin{minipage}{0.49\textwidth}
        \centering
           \includegraphics[width=0.95\columnwidth]{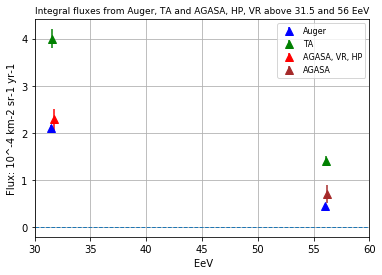}
    \end{minipage}
    \begin{minipage}{0.49\textwidth}
        \centering
        \footnotesize
        \begin{tabular}{l c c c}
            \textbf{Observatory} &  \textbf{$A\Omega t$} & \textbf{N} & \textbf{Flux ($I \geq 31.5$\,EeV)}\\
             & $\mathrm{km}^2\,\mathrm{sr}\,\mathrm{yr}$ & & $\mathrm{km}^{-2}\,\mathrm{sr}^{-1}\,\mathrm{yr}^{-1}$\\
            \hline
            Auger & 60,400 & 1262 & 0.0210$\pm$0.0006 \\
            TA & 13,565 & 538 & 0.040$\pm$0.002 \\
            AGASA & 1617 & 35 & 0.022$\pm$0.004 \\
              + VR and HP & 1927 & 46 & 0.023$\pm$0.003 \\
             & & & \\
            \textbf{Observatory} &  \textbf{$A\Omega t$} & \textbf{N} & \textbf{Flux ($I \geq 56$\,EeV)}\\
            \hline
            Auger & 60,400 & 270 & 0.0045$\pm$0.0003 \\
            TA & 13,565 & 189 & 0.014$\pm$0.01 \\
            AGASA & 1617 & 11 & 0.007$\pm$0.002 \\         
        \end{tabular}
    \end{minipage}
    \caption{Comparison of the integral fluxes above 31.5 and 56 EeV measured by the AGASA (after adjustment), Auger and Telescope Array Collaborations. Eleven events from Volcano Ranch and Haverah Park have been included at the lower energy.}
    \label{fig:integral_fluxes}
\end{figure}

In Figure~\ref{fig:AGASA_TA_residuals}, the complete AGASA and Telescope Array (TA) data sets are displayed in residual plots.
The TA measurements begin to fall outside the Auger 14\% bounds above $\log E = 19.5$, whereas the AGASA data lie within these bounds except for the last three points (the 11 highest energy events).
For all the data, the respective reduced $\chi^2$ values are 4.35 and 7.01 but the fit at the highest energies, the crucial region, is clearly better for the AGASA data.
Data from Volcano Ranch and Haverah Park are useful in the lower energy range and have been added to those from AGASA.
A comparison of the integral flux measurements from the two detectors above 31.5 and 56 EeV is shown in Figure~\ref{fig:integral_fluxes}.
It is evident that the historic data are in better agreement with the Auger measurement than with that from the Telescope Array.

\begin{figure}[t]
    \centering
    \includegraphics[width=0.49\columnwidth]{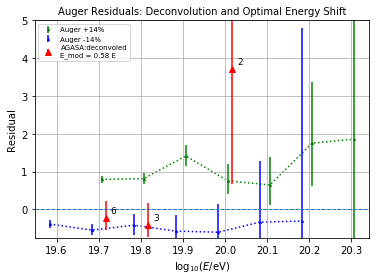}
    \caption{Residual plot for the 11 highest-energy events reported by AGASA. The reduced value $\chi^2$ is 0.61.}
    \label{fig:AGASA_residuals_highenergy}
\end{figure}

In Figure~\ref{fig:AGASA_residuals_highenergy}, a residual plot is shown for the 11 highest energy events from AGASA.
The shift in energy needed to get agreement with the Auger flux is 0.58.
This confirms the results that an energy independent correction factor $k$ could not be applied and that, depending on the energy range, AGASA energy estimate should be reduced by $\sim$0.8 to $\sim$0.6.
This is an extremely large adjustment that might imply that the highest-energy events are created by trans-iron nuclei or that there are exceptionally energetic events coming from sources unique to the Northern Hemisphere.
However, there are considerable uncertainties about the energy assessments made for four of the highest-energy events for which details are available.

At the time for the AGASA project, the computing power available limited the range of simulations that were possible and the relationship between $S(600)$, the signal at 600\,m adopted to characterise the shower size parameter, and the primary energy was computed only for vertical events.
Values of $S(600)$ measured at other zenith angles were adjusted to a vertical value using an attenuation length which cannot be measured with precision at the greatest energies because of the small number of events.
Using EPOS-LHC and QGSJetII-04, we have evaluated the energies for the four highest-energy events.
Results are shown in Table~\ref{tab:AGASA_highestevents}.
\begin{table}[t]
    \centering
    \small
    \begin{tabular}{l|c|c|c|c|c|c}
        \multirow{2}{*}{Event} & \multirow{2}{*}{$E$ / EeV} & \multirow{2}{*}{$\theta$ / $^\circ$} & \multicolumn{2}{c}{EPOS-LHC} & \multicolumn{2}{|c}{QGSJetII.04} \\
         & & & Proton & Iron & Proton & Iron \\
        \hline
        \hline
        25400-0296~\cite{Hayashida_1994} & 260\footnotemark & 22.9 & 0.79 & 0.70 & 0.82 & 0.76 \\
        unknown id~\cite{Sakaki_2001} & 280 & 37 & 0.49 & 0.49 & 0.51 & 0.54 \\
        01606-0578~\cite{Doi_1997} & 150 & 44.2 & 0.42 & 0.48 & 0.48 & 0.55 \\
        00123-3997~\cite{Doi_1997} & 144 & 14.1 & 0.76 & 0.65 & 0.80 & 0.70 \\
    \end{tabular}
    \caption{Estimates of the energies of the four high energy AGASA events for which details are available.}
    \label{tab:AGASA_highestevents}
\end{table}
\footnotetext[1]{The upper-limit of the estimate of the energy has been used to simulate this event.}
The adjustment in energy using EPOS-LHC, averaging over protons and iron for the two near-vertical events is $\sim$73\%, while for the events at 37$^\circ$ and 44.2$^\circ$ the corresponding average reduction is $\sim$47\%.
These differences are confirmed using QGSJetII-04 as the model for the hadronic interactions.
Although the hadronic model is known to be inadequate (see discussion of Figure 1), we consider these evaluations of the energies to be the best currently possible.

\begin{figure}[t]
    \centering
    \includegraphics[width=0.49\columnwidth]{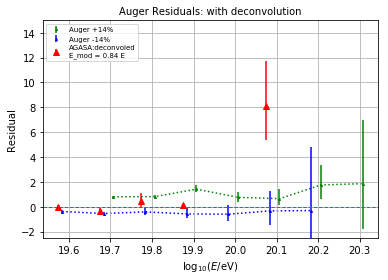}
    \includegraphics[width=0.49\columnwidth]{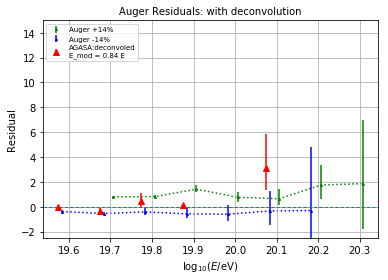}
    \caption{Residual plots for the highest energy events reported by AGASA: see text for details. The reduced values of $\chi^2$ are 2.16 and 1.25 respectively.}
    \label{fig:AGASA_teshima_correction}
\end{figure}

In an unpublished presentation [15] given around 2006, Teshima made a general statement about an energy reduction applicable to the 11 highest energy events.
He commented that there should be a downward shift of $\sim$15\% and that the 11 events are reduced to 5 to 6 events.
No details on an event-by-event basis were given.
In Figure~\ref{fig:AGASA_teshima_correction}, the fluxes for the highest energy AGASA events are shown with the final point calculated for an interval in energy, $\Delta\log E$ (EeV), from 20 – 20.4 (100 – 250 EeV).
Following Teshima (right-hand plot), only five events above 100 EeV remain.
Thus, no exotic explanation is necessary to understand the highest energy events.

\begin{figure}[t]
    \centering
    \includegraphics[width=0.6\columnwidth]{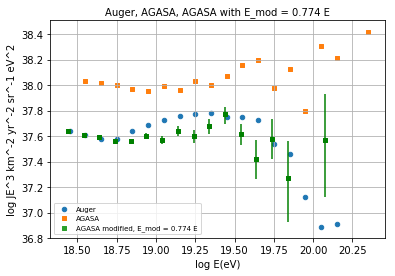}
    \caption{A comparison of the revised AGASA energy spectrum with that from the Auger Observatory.  The original AGASA data are in orange with the revised values in green.  Further details are given in the text.}
    \label{fig:AGASA_all_corrections}
\end{figure}

In Figure~\ref{fig:AGASA_all_corrections}, a plot of $\log E^3 J$ vs $\log E$, as an update of Figure~\ref{fig:corrected_spectra}, is shown for AGASA applying the different corrections discussed in this paper.
The agreement between the AGASA spectrum and that measured by the Auger Collaboration is greatly improved.
Note that no attempt has been made to take into account the complex change in mass composition across the energy range~\cite{AugerXmax_PRD2026} with which the energy adjustment, $k$, will change as the energy increases.

\section{Conclusion}
Assuming that the fluxes reported from the Auger Observatory are accurate, and that the fluxes at energies lower than the ankle are the same in both hemispheres, we have shown that the fluxes from historic measurements above 31 and 56 EeV are in better agreement with the results reported by the Auger Collaboration than with what has been presented the Telescope Array.
The fluxes from these historic measurements, all from the Northern Hemisphere, are a factor of two lower than those reported by the Telescope Array Collaboration.
This conclusion is independent of any assumptions about hadronic interaction models.
We conclude, therefore, that the Telescope Array Collaboration might wish to consider using a different model for hadronic interactions and adopting a heavier mass composition than the all-proton assumption currently chosen.
We also suggest that more attention should be given to matching model predictions with direct measurements of the energy spectrum at energies where mass estimates are most reliable.

\acknowledgments
We thank Jonathan Biteau and Olivier Deligny for their help and guidance of QL in the production of the simulations used to tackle this problem.

\end{document}